\documentclass[aps,twocolumn,pra,showpacs,superscriptaddress]{revtex4}
\usepackage{amssymb}
\usepackage{amsmath}
\usepackage{graphicx}
\usepackage{epsfig}
\usepackage{subfigure}
\usepackage{amsfonts}

\input{tcilatex}

\begin{document}

\title{Multi-Stability of Electromagnetically Induced Transparency in
Atom-Assisted Optomechanical Cavities}
\author{Yue Chang}
\affiliation{Institute of Theoretical Physics, Chinese Academy of Sciences, Beijing,
100190, China}
\author{T. Shi}
\affiliation{Institute of Theoretical Physics, Chinese Academy of Sciences, Beijing,
100190, China}
\author{Yu-xi Liu}
\affiliation{Advanced Science Institute, The Institute of Physical and Chemical Research
(RIKEN), Wako-shi 351-0198, Japan}
\affiliation{CREST, Japan Science and Technology Agency (JST), Kawaguchi, Saitama
332-0012, Japan}
\author{C. P. Sun}
\affiliation{Institute of Theoretical Physics, Chinese Academy of Sciences, Beijing,
100190, China}
\author{Franco Nori}
\affiliation{Advanced Science Institute, The Institute of Physical and Chemical Research
(RIKEN), Wako-shi 351-0198, Japan}
\affiliation{CREST, Japan Science and Technology Agency (JST), Kawaguchi, Saitama
332-0012, Japan}
\affiliation{Center for Theoretical Physics, Physics Department, Center for the Study of
Complex Systems, The University of Michigan, Ann Arbor, MI 48109-1040, USA.}

\begin{abstract}
We study how an oscillating mirror affects the electromagnetically induced
transparency (EIT) of an atomic ensemble, which is confined in a gas cell
placed inside a micro-cavity with an oscillating mirror in one end. The
oscillating mirror is modeled as a quantum mechanical harmonic oscillator.
The cavity field acts as a probe light of the EIT system and also produces a
light pressure on the oscillating mirror. The back-action from the mirror to
the cavity field results in several (from one to five) steady-states for
this atom-assisted optomechanical cavity, producing a complex structure in
its EIT. We calculate the susceptibility with respect to the few (from one
to three) stable solutions found here for the equilibrium positions of the
oscillating mirror. We find that the EIT of the atomic ensemble can be
significantly changed by the oscillating mirror, and also that the various
steady states of the mirror have different effects on the EIT.
\end{abstract}

\maketitle

\section{Introduction}

Fast developments are now occurring in studies at the interface between
different kinds of physical systems. Examples of these include: couplings
between light and nano/micro-mechanical systems (e.g.,~\cite%
{mancini,meystre,vitali,penrose,tombesi2,tombesi3,meystre2,meystre3,meystre4,meystre5}%
), called optomechanical systems; interactions between
superconducting artificial atoms (e.g., charge or flux qubit) and
transmission line resonators~(e.g.,
\cite{nori,walraff,nori2,nori3,nori4,Delft}), and so on. These
studies are partly motivated by possible physical implementations
of quantum information processing, to explore potential future
devices, and to study the interesting physics in these hybrid
structures.

In optomechanical systems (e.g.,~\cite%
{mancini,meystre,vitali,penrose,tombesi2,tombesi3,meystre2,meystre3,meystre5}%
), the radiation pressure acts on the oscillating mirror and induces the
interaction between the mechanical system and the optical field. The
back-action from the mirror to the cavity field \cite{berman} can result in
several (about 5) steady-state solutions for the equilibrium position of the
mirror. It has been proved~\cite{mancini} that this system experiences
bistability in some parameter region. It has also been shown that light
pressure can be used to realize entanglement between the cavity field and a
microscopic object (e.g., a moveable mirror)~\cite%
{mancini,meystre,vitali,penrose}, and can also cool down the mirror~\cite%
{tombesi2,tombesi3,meystre2,meystre3}. Furthermore, Refs.~\cite{tombesi,ian}
studied optomechanical cavities containing a two-level atomic ensemble. The
atomic ensemble effectively enhances the radiation pressure of the cavity
field on the oscillating mirror, producing a cavity-atom-mirror entanglement~%
\cite{tombesi,ian}. In this paper we consider this atom-optical system with
a more complex atomic ensemble, which can enable quantum interference, e.g.,
electromagnetically induced transparency.

Electromagnetically induced transparency (EIT) is a remarkable quantum
interference phenomenon, which enables the active control of light
propagation in a coherent medium~\cite{harris,aaron,hau,lukin}. Usually, the
basic population transfer configuration for the atoms in EIT is of $\Lambda $%
-type, where the two transitions from a common upper energy level to two
lower energy levels are induced by two different optical fields (e.g.,
classical control field and probe quantized field)~\cite%
{scully1,liyong,lukin1}, respectively. One is a strong field, and the other
is a weak one. The stronger field can effectively modify the susceptibility
of the medium so that the weak one (as a probe signal) can pass through the
medium transparently at the two-photon-resonance~\cite%
{scully,deng,kozuma,liyong1,he}. Recently, it has been shown that this
effect can also work well at the single photon level for the probe light~%
\cite{gong}, and thus the weak field must be treated quantum mechanically.
In the quantum approach, a dark state with dressed photons can be invoked to
store quantum information of photons on the atomic ensemble as quantum
memory~\cite{lukin2,lukin3}. These quantum manipulations at the single
photon level require frequency-matching with extremely high-precision for
one- or two- photon resonances. When the quantum field is provided by a
micro-cavity with a one-end oscillating mirror, the oscillation of the
mirror might affect such a precise frequency-matching condition and thus
affect the EIT.

Motivated by these works, here we study how the oscillating mirror changes
the properties of the EIT. We will study the atom-assisted optomechanical
micro-cavity, through which we explore the possibility to interface other
systems, via some physical mechanism, such as EIT. Here, the atomic ensemble
for each atom with $\Lambda $-type transitions is placed inside a cavity
with a one-end oscillating mirror. Due to the mirror's oscillation, the
susceptibility of this atomic medium displays a multi-stability
corresponding to the multi-equilibrium positions of the mirror. Another
prediction of our study is that the mirror's oscillation significantly
alters the properties of both the real and imaginary parts of the
susceptibility. We also investigate in detail how the different steady
states of the mirror influence the dispersion relations and absorption
properties of the light.

The paper is organized as follows: in Sec.~II, we introduce the model for
the optomechanical system with EIT; in Sec.~III, we present the
Heisenberg-Langevin equations for this system and obtain several (from one
to five, depending on the system parameters) steady-state solutions for the
position of the oscillating mirror; in Sec.~IV, we study the EIT with
susceptibilities for different (from one to two) equilibrium positions of
the mirror; in Sec.~V, we summarize our results.

\section{atom-assisted optomechanical system}

\begin{figure}[tbp]
\includegraphics[bb=134 306 426 633, width=8 cm, clip]{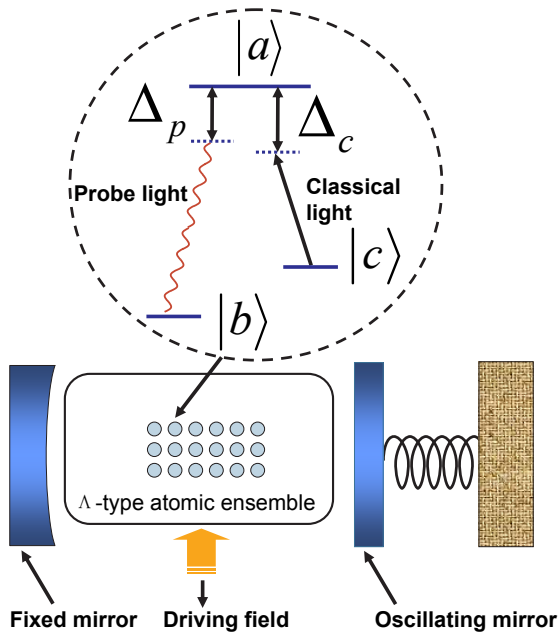}
\caption{(Color online) Schematic diagram for the atom-assisted
optomechanical system considered here. There are three main components: (i)
an optical cavity with one fixed mirror; (ii) another oscillating mirror,
which is modeled as a quantum-mechanical harmonic oscillator (denoted by a
black spring); and (iii) an ensemble of identical three-level atoms, which
are confined inside the cavity. Each atom is assumed to have $\Lambda$-type
transitions. Here, $\Delta _{p}=\protect\omega _{a}-\protect\omega _{0}+%
\protect\omega _{0}\left\langle x\right\rangle /l$ and $\Delta _{c}=\protect%
\omega _{a}-\protect\omega _{c}$.}
\label{fig1}
\end{figure}

As shown in Fig.~\ref{fig1}, we consider an ensemble of $N$ identical
three-level atoms, which are confined inside an optical cavity with a
one-end oscillating mirror. The excited, meta-stable, and ground states of
the $i$th atom are denoted by $\left\vert a\right\rangle _{i}$, $\left\vert
c\right\rangle _{i}$, and $\left\vert b\right\rangle _{i}$. Each atom is
assumed to have $\Lambda $-type transitions. That is, for the $i$th atom, a
classical light field with frequency $\nu $ induces a transition between $%
\left\vert a\right\rangle _{i}$ and $\left\vert c\right\rangle _{i}$, which
is often used as a control field. The quantized cavity field, with frequency
$\omega _{0}$ when the oscillating mirror is fixed, induces the transition
between $\left\vert a\right\rangle _{i}$ and $\left\vert b\right\rangle _{i}$%
. We assume that this cavity field acts as a probe field. Here, the
transition between $\left\vert b\right\rangle _{i}$ and $\left\vert
c\right\rangle _{i}$ is assumed to be forbidden. The oscillating mirror is
modeled as a quantum-mechanical harmonic oscillator with frequency $\omega
_{M}$ and the mass $M$. This harmonic oscillator can also be considered as a
spring with an elastic coefficient $M\omega _{M}^{2}$.

Based on the above considerations, and using $\hbar =1$, the Hamiltonian of
the total system%
\begin{equation}
H=H_{C}+H_{M}+H_{A}+H_{M\text{-}C}+H_{A\text{-}L}  \label{eq:1}
\end{equation}%
has five terms corresponding to the cavity ($H_{C}$), the oscillating mirror
($H_{M}$) of mass $M$, the atom gas ($H_{A}$), the mirror-cavity ($H_{M\text{%
-}C}$), and the atom-light term ($H_{A\text{-}L}$). Explicitly, these are
described below: (i) the free Hamiltonian
\begin{equation}
H_{C}=\omega _{0}a^{\dag }a,  \label{2}
\end{equation}%
of the single-mode cavity field with the annihilation and creation operators
$a$ and $a^{\dag }$; this term (and $\omega _{0}$) refers to a cavity with
two fixed mirrors; (ii) the free Hamiltonian
\begin{equation}
H_{M}=\frac{p^{2}}{2M}+\frac{1}{2}M\omega _{M}^{2}x^{2},
\end{equation}%
of the oscillating mirror, where $p$ is the momentum of the oscillating
mirror with a small displacement $x$; (iii) the free Hamiltonian
\begin{equation}
H_{A}=\sum\limits_{i=1}^{N}(\omega _{a}\sigma _{aa}^{\left( i\right)
}+\omega _{c}\sigma _{cc}^{\left( i\right) }),
\end{equation}%
of the $N$ three-level $\Lambda $-atoms with the operators $\sigma _{\alpha
\alpha }^{\left( i\right) }=|\alpha \rangle _{ii}\langle \alpha |$ and $%
\alpha =a,\,c$; here, $\omega _{a}$ ($\omega _{c}$) is the energy level
spacing between $\left\vert a\right\rangle _{i}$ and $\left\vert
b\right\rangle _{i}$ ($\left\vert c\right\rangle _{i}$ and $\left\vert
b\right\rangle _{i}$) for the $i$th atom, and we have assumed the ground
state $|b\rangle _{i}$ as an energy reference point; (iv) the interaction
Hamiltonian
\begin{equation}
H_{M\text{-}C}=-\frac{\omega _{0}}{l}\,x\,a^{\dag }a,
\end{equation}%
between the cavity field and the oscillating mirror~\cite{mancini}, presents
a radiation pressure on the mirror due to the small change $x$ of the cavity
length when the mirror oscillates~\cite{tombesi2}, where $l$ is the cavity
length when the mirror is at its equilibrium position; (v) the interaction
Hamiltonian
\begin{equation}
H_{A\text{-}L}=\sum\limits_{i=1}^{N}(\Omega \,e^{-i\nu t}\,\sigma
_{ac}^{\left( i\right) }+g\,a\,\sigma _{ab}^{\left( i\right) }+h.c.),
\label{eq:6}
\end{equation}%
between the three-level atoms and the classical as well as the quantized
fields. In Eq.~(\ref{eq:6}), $\Omega $ is the Rabi frequency associated with
the coupling between the classical field and the three-level atoms. The
frequency $\nu $ is assumed to satisfy the condition $\nu =\omega
_{a}-\omega _{c}-\Delta _{c}$. Here, $\Delta _{c}$ is the detuning between
the frequency of the classical control field and the transition frequency
from the energy level $|a\rangle _{i}$ to the energy level $|c\rangle _{i}$
for the $i$th atom.

The parameter
\begin{equation}
g=-\mu \sqrt{\omega _{0}/2V\epsilon _{0}}
\end{equation}
in (\ref{eq:6}) describes the coupling between the quantized cavity field
and the three-level atoms, where $\mu $ is the electric-dipole transition
matrix element between levels $|a\rangle $ and $|b\rangle $, $V$ describes
the volume of the cavity, and $\epsilon _{0}$ is the permittivity of the
vacuum. We note that the effect of the oscillating mirror on the coupling
between the atoms and the quantized cavity field~\cite{chang} has been
neglected when the Hamiltonian in Eq.~(\ref{eq:1}) was derived, because we
do not consider the strong coupling between the quantized field and the
atoms.

In the derivation of the Hamiltonian in Eq.~(\ref{eq:1}), we have assumed
that the linear size of the atomic ensemble is much smaller than the
wavelengths of the light fields. In this case, the couplings between the
atoms and the light fields are homogeneous, and we can define the collective
operators of the atomic ensemble as in Refs.~\cite{liyong,liyong1}
\begin{eqnarray}
S &=&\sum\limits_{i=1}^{N}\sigma _{aa}^{\left( i\right) }\text{,}%
\;\;\;\;A^{\dag }=\frac{1}{\sqrt{N}}\sum\limits_{i=1}^{N}\sigma
_{ab}^{\left( i\right) }\text{,}  \notag \\
T_{+} &=&(T_{-})^{\dag }=\sum\limits_{i=1}^{N}\sigma _{ac}^{\left( i\right) }%
\text{,}\;\;\;\;C=\frac{1}{\sqrt{N}}\sum\limits_{i=1}^{N}\sigma
_{bc}^{\left( i\right) }\text{.}  \label{eq:7}
\end{eqnarray}%
Together with Eq.~(\ref{eq:7}), the interaction Hamiltonian $H_{A\text{-}C}$
in Eq.~(\ref{eq:6}) can be rewritten as
\begin{equation}
H_{A\text{-}L}=\Omega \,e^{-i\nu t}\,T_{+}+g\sqrt{N}\,a\,A^{\dag }+\text{H.c.%
}\,\,.
\end{equation}%
We assume that the number $N$ of atoms is large enough so that the
collective operators in Eq.~(\ref{eq:7}) satisfy the communication relations
as in Refs.~\cite{liyong,liyong1,ian}
\begin{subequations}
\begin{equation}
\left[ C^{\dag },S\right] =0\text{, }\;\;\;\left[ A,S\right] =A,  \label{6a}
\end{equation}%
and%
\begin{equation}
\left[ T_{-},C^{\dag }\right] =0\text{, }\;\;\;\left[ T_{+},C^{\dag }\right]
=A^{\dag },  \label{6b}
\end{equation}%
when most atoms are in their ground states. Equations~(\ref{6a}) and (\ref%
{6b}) present a dynamical symmetry in our system described by the
semidirect-product algebra containing the algebra SU(2) with its generators $%
T_{\pm }$. It is easy to prove, as in Ref.~\cite{ian}, that the collective
operators and the communication relations can also be given in a similar way
as in Eqs.~(\ref{eq:6}), (\ref{6a}) and (\ref{6b}) for the case when the
couplings between different atoms and the light fields are inhomogeneous.
Therefore, our study here can be generalized to the case of inhomogeneous
couplings.

\section{Heisenberg-Langevin equations and multi-stability}

\subsection{Steady-state positions of the moveable mirror: analytical results%
}

Using the commutation relations in Eqs.~(\ref{6a})-(\ref{6b}), we can write
the Heisenberg-Langevin equations of motion as
\end{subequations}
\begin{subequations}
\begin{equation}
\partial _{t}x=\frac{p}{M},
\end{equation}%
\begin{equation}
\partial _{t}p=-\frac{\gamma _{M}}{2M}p+\frac{\omega _{0}}{l}a^{\dag
}a-M\omega _{M}^{2}\,x-\sqrt{\gamma _{M}}\epsilon _{\mathrm{in}}\left(
t\right) ,  \label{7a}
\end{equation}%
\begin{equation}
\partial _{t}a=-\frac{\gamma _{0}}{2}a-i\omega _{0}\left( 1-\frac{x}{l}%
\right) a-ig\sqrt{N}\,A+\sqrt{\gamma _{0}}a_{\mathrm{in}}\left( t\right) ,
\end{equation}%
\begin{equation}
\partial _{t}A=-\gamma _{1}A-i\omega _{a}A-i\Omega \,e^{-i\nu t}\,C-ig\sqrt{N%
}\,a+f_{1}(t),
\end{equation}%
\begin{equation}
\partial _{t}C=-\gamma _{2}C-i\omega _{c}C-i\Omega \,e^{i\nu t}\,A+f_{2}(t).
\label{7d}
\end{equation}%
Here, $\partial _{t}$ denotes a time derivative. We have phenomenologically
introduced the damping $\gamma _{M}$ of the oscillating mirror, the decay
rates $\gamma _{0}$ for the cavity field and $\gamma _{1}$ ($\gamma _{2}$)
for the decay from $\left\vert a\right\rangle $ to $\left\vert
b\right\rangle $ ($\left\vert c\right\rangle $ to $\left\vert b\right\rangle
$), respectively. We also assume that the quantum fluctuations of the cavity
field, mirror, and the atoms satisfy the conditions
\end{subequations}
\begin{equation}
\left\langle \epsilon _{\mathrm{in}}\left( t\right) \right\rangle
=\left\langle f_{1}\left( t\right) \right\rangle =\left\langle f_{2}\left(
t\right) \right\rangle =0
\end{equation}%
and%
\begin{equation}
\left\langle a_{\mathrm{in}}\left( t\right) \right\rangle =\alpha _{\mathrm{%
in}}\left( t\right) .
\end{equation}%
Here, $\alpha _{\mathrm{in}}\left( t\right) $ can be understood as an input
driving field. That is, $a_{\mathrm{in}}\left( t\right) $ can be rewritten
as
\begin{equation*}
a_{\mathrm{in}}\left( t\right) =\alpha _{\mathrm{in}}\left( t\right) +\delta
a_{\mathrm{in}}\left( t\right) ,
\end{equation*}%
where the quantum fluctuation of the input field $\delta a_{\mathrm{in}%
}\left( t\right) $ satisfies $\left\langle \delta
a_{\mathrm{in}}\left( t\right) \right\rangle =0$.

To obtain the steady-state solutions, let us first remove the fast varying
factors by the following rotating transformations
\begin{subequations}
\begin{eqnarray}
a &=&\tilde{a}\exp (-i\omega _{L}t), \\
A &=&\tilde{A}\exp (-i\omega _{L}t), \\
C &=&\tilde{C}\exp \left[ i\left( \Delta _{p}-\Delta _{c}-\omega _{c}\right)
t\right] ,
\end{eqnarray}%
and
\begin{eqnarray}
a_{\mathrm{in}}\left( t\right) &=&\tilde{a}_{\mathrm{in}}\left( t\right)
\exp (-i\omega _{L}t) \\
&=&[\tilde{\alpha}_{\mathrm{in}}(t)+\delta
\tilde{a}_{\mathrm{in}}\left( t\right) ]\exp (-i\omega _{L}t),
\end{eqnarray}%
where the detuning $\Delta _{p}$ between the transition frequency $\omega
_{a}$ of the atom, between the energy levels $|a\rangle $ and $|b\rangle $,
and the effective frequency $\omega _{L}$ of the cavity field, is given by
\end{subequations}
\begin{equation}
\Delta _{p}=\omega _{a}-\omega _{L},  \label{eq:14}
\end{equation}%
with the effective cavity frequency
\begin{equation}
\omega _{L}=\omega _{0}-\frac{\omega _{0}}{l}\left\langle x\right\rangle .
\label{eq:15}
\end{equation}%
Here, $\left\langle x\right\rangle $ denotes the mean value of $x$.
Equation~(\ref{eq:15}) shows that the effective frequency $\omega _{L}$ of
the cavity can be changed by the oscillating mirror. When the oscillating
mirror has zero displacement, then $\omega _{L}$ equals to $\omega _{0}$.

In the rotating frame, the Heisenberg-Langevin equations in Eqs.~(\ref{7a})-(%
\ref{7d}) become

\begin{subequations}
\begin{equation}
\partial _{t}p=-\frac{\gamma _{M}}{2M}p+\frac{\omega _{0}}{l}\tilde{a}^{\dag
}\tilde{a}-M\omega _{M}^{2}\,x-\sqrt{\gamma _{M}}\epsilon _{\mathrm{in}%
}\left( t\right) ,  \label{m}
\end{equation}%
\begin{equation}
\partial _{t}\tilde{A}=-(\gamma _{1}+i\Delta _{p})\tilde{A}-i\Omega \tilde{C}%
-ig\sqrt{N}\,\tilde{a}+\tilde{f}_{1}(t),
\end{equation}%
\begin{equation}
\partial _{t}\tilde{C}=-\left[ \gamma _{2}+i\left( \Delta _{p}-\Delta
_{c}\right) \right] \tilde{C}-i\Omega \,\tilde{A}+\tilde{f}_{2}(t),
\end{equation}%
\begin{equation}
\partial _{t}\tilde{a}=-\left[ \frac{\gamma _{0}}{2}-i\frac{\omega _{0}}{l}%
\left( \left\langle x\right\rangle -x\right) \right] a-ig\sqrt{N}\,\tilde{A}+%
\sqrt{\gamma _{0}}\tilde{a}_{\mathrm{in}}\left( t\right) ,  \label{n}
\end{equation}%
where the fluctuation forces are
\end{subequations}
\begin{subequations}
\begin{equation}
\tilde{f}_{1}(t)=f_{1}(t)\exp \left[ i\omega _{0}\left( 1+\frac{\left\langle
x\right\rangle }{l}\right) t\right] ,
\end{equation}%
and%
\begin{equation}
\tilde{f}_{2}(t)=f_{2}(t)\exp \left[ -i\left( \Delta _{p}-\Delta _{c}-\omega
_{c}\right) t\right] .
\end{equation}%
We are interested in the steady-state regime. Let us first assume that all
the time derivatives of the mean values for the operators in Eqs.~(\ref{m})-(%
\ref{n}) are equal to zero; then we obtain the steady-state equations
\end{subequations}
\begin{subequations}
\begin{equation}
\frac{\omega _{0}}{l}\left\langle \tilde{a}^{\dag }\right\rangle
_{s}\left\langle \tilde{a}\right\rangle _{s}-M\omega _{M}^{2}\left\langle
x\right\rangle _{s}=0,  \label{d}
\end{equation}%
\begin{equation}
-\frac{\gamma _{0}}{2}\left\langle \tilde{a}\right\rangle _{s}-ig\sqrt{N}%
\left\langle \tilde{A}\right\rangle _{s}+\sqrt{\gamma _{0}}\tilde{\alpha}_{%
\mathrm{in}}=0,  \label{a}
\end{equation}%
\begin{equation}
-\left( \gamma _{1}+i\Delta _{p,s}\right) \left\langle \tilde{A}%
\right\rangle _{s}-i\Omega \left\langle \tilde{C}\right\rangle _{s}-ig\sqrt{N%
}\left\langle \tilde{a}\right\rangle _{s}=0,
\end{equation}%
\begin{equation}
-\left[ \gamma _{2}+i\left( \Delta _{p,s}-\Delta _{c}\right) \right]
\left\langle \tilde{C}\right\rangle _{s}-i\Omega \,\left\langle \tilde{A}%
\right\rangle _{s}=0.  \label{b}
\end{equation}%
Here, $\left\langle O\right\rangle _{s}$ ($O$ represents the operator in the
above steady-state equations) is the mean value of the operator $O$ under
the steady state. The parameter $\Delta _{p,s}$ is the detuning described in
Eq.~(\ref{eq:14}) when the system reaches steady-state. In Eqs.~(\ref{d})
and (\ref{a}), the correlations $\langle \tilde{a}^{\dag }\tilde{a}\rangle
_{s}$ and $\langle x\tilde{a}\rangle _{s}$ have been approximately replaced
by $\langle \tilde{a}^{\dag }\rangle _{s}\langle \tilde{a}\rangle _{s}$ and $%
\langle x\rangle _{s}\langle \tilde{a}\rangle _{s}$, respectively. These
approximations indicate that the correlations between the fluctuations near
the steady states are very small compared to the corresponding mean values
in the steady states. This can be quantitatively shown as~\cite{le}
\end{subequations}
\begin{equation}
\frac{\left\langle (\delta \tilde{a}^{\dag })\,(\delta \tilde{a}%
)\right\rangle _{s}}{\left\langle \tilde{a}^{\dag }\right\rangle
_{s}\left\langle \tilde{a}\right\rangle _{s}}\ll 1\text{, }\;\;\;\frac{%
\left\langle (\delta x)\,(\delta \tilde{a})\right\rangle _{s}}{%
\left\langle x\right\rangle _{s}\left\langle \tilde{a}\right\rangle _{s}}\ll
1.
\end{equation}

From Eqs.~(\ref{a})-(\ref{b}), when the system reaches the steady-state, the
mean values $\langle \tilde{A}\rangle _{s}$ and $\left\langle \tilde{a}%
\right\rangle _{s}$ can be easily derived as
\begin{equation}
\left\langle \tilde{A}\right\rangle _{s}=\frac{-ig\sqrt{N\gamma _{0}}\,%
\tilde{\alpha}_{\mathrm{in}}\,\tilde{\Omega}\!\left( \Delta _{p,s}\right) }{%
G\!\left( \Delta _{p,s}\right) \,\tilde{\Omega}\!\left( \Delta _{p,s}\right)
+\frac{\gamma _{0}}{2}\Omega ^{2}},  \label{a1}
\end{equation}%
and
\begin{equation}
\left\langle \tilde{a}\right\rangle _{s}=\frac{2\tilde{\alpha}_{\mathrm{in}}%
}{\sqrt{\gamma _{0}}}\left[ 1-\frac{g^{2}N\,\tilde{\Omega}\!\left( \Delta
_{p,s}\right) }{G\!\left( \Delta _{p,s}\right) \,\tilde{\Omega}\!\left(
\Delta _{p,s}\right) +\frac{\gamma _{0}}{2}\Omega ^{2}}\right] ,  \label{c}
\end{equation}%
with the functions
\begin{equation}
\tilde{\Omega}\!\left( \Delta _{p,s}\right) =\gamma _{2}+i\left( \Delta
_{p,s}-\Delta _{c}\right)
\end{equation}%
and
\begin{equation}
G\!\left( \Delta _{p,s}\right) =g^{2}N+\frac{1}{2}\gamma _{0}\left( \gamma
_{1}+i\Delta _{p,s}\right) .
\end{equation}

Using Eq.~(\ref{d}) and Eq.~(\ref{c}), we can self -consistently derive a
nonlinear implicit equation for $\Delta _{p,s}$ as below
\begin{equation}
\left\vert 1-\frac{g^{2}N\,\tilde{\Omega}\!\left( \Delta _{p,s}\right) }{%
G\!\left( \Delta _{p,s}\right) \,\tilde{\Omega}\!\left( \Delta _{p,s}\right)
+\frac{\gamma _{0}}{2}\Omega ^{2}}\right\vert ^{2}=\frac{\gamma _{0}\kappa }{%
4\tilde{\alpha}_{\mathrm{in}}^{2}\omega _{0}^{2}}\left( \Delta _{p,s}-\Delta
_{0}\right) ,  \label{e}
\end{equation}%
where the elastic constant (which has units of energy) of the spring
attached to the moveable mirror is%
\begin{equation}
\kappa =M\omega _{M}^{2}l^{2},  \label{eq:25}
\end{equation}%
and
\begin{equation}
\Delta _{0}=\omega _{a}-\omega _{0}.
\end{equation}%
for the detuning $\Delta _{0}$ between $\omega _{a}$ (the highest frequency
of the $\Lambda $-atom) and $\omega _{0}$ (the cavity frequency in Eq. (\ref%
{2}) when the two mirrors are fixed). Equation~(\ref{e}) is a fifth-power
implicit equation for $\Delta _{p,s}$. Thus the system may have several
solutions (i.e., multi-stability in some parameter regions, corresponding to
several steady-state positions of the mirror). These mirror positions are
determined by
\begin{equation}
\left\langle x\right\rangle _{s}=\frac{\omega _{0}\left\langle \tilde{a}%
^{\dag }\right\rangle _{s}\,\left\langle \tilde{a}\right\rangle _{s}}{%
M\omega _{M}^{2}l}.
\end{equation}

\begin{figure}[tbp]
\includegraphics[bb=34 169 500 667, width=9 cm, height=11 cm, clip]{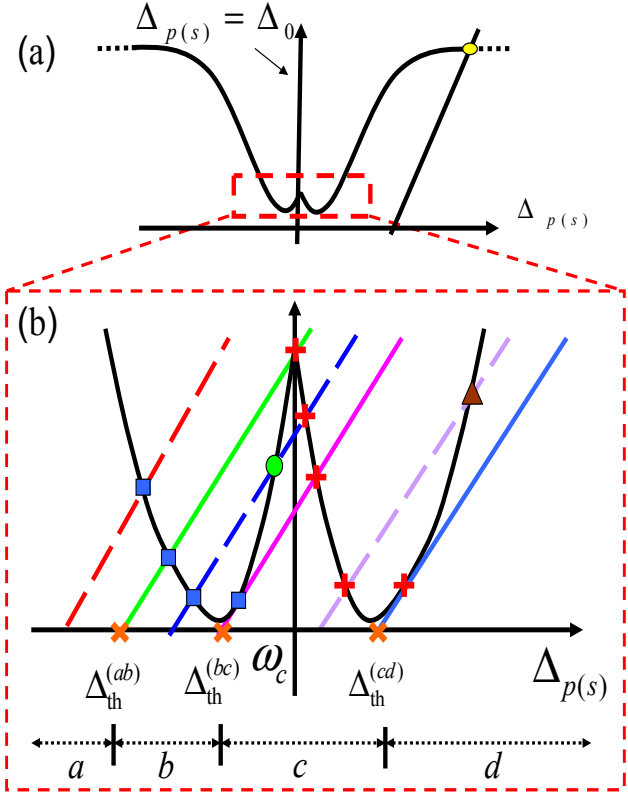}
\caption{(Color online) Schematic diagram for the functions of the left $Y_{%
\mathrm{L}}\!\left( \Delta _{p,s}\right)$ and the right $Y_{\mathrm{R}%
}\!\left( \Delta _{p,s}\right)$ hand sides of Eq.~(\protect\ref{e}) versus $%
\Delta_{p,s}$ in (a) for a large value of $\Delta_{0}$ and (b) for $%
\Delta_{0}$ in the region of (a) indicated by the red dashed box. In (b),
the brown triangle, red crosses, green dot and blue squares denote four
different steady state solutions of the self-consistent equation~(\protect
\ref{e}). The brown triangle (far right) denotes the solution $\Delta
_{p,s}^{(1)}$, the four blue squares (top left) denote the solution $\Delta
_{p,s}^{(2)}$, the red dots on the curve denotes the solution $\Delta
_{p,s}^{(3)}$, and the green dot denotes the solution $\Delta _{p,s}^{(4)}$.
The solution $\Delta _{p,s}^{(1)}$ is not shown in the regions c and d.
Here, $\Delta_{\mathrm{th}}^{(i)}$ ($i=ab,bc,cd$) denote the threshold
values of $\Delta_{0}$ that separate regions with different number of
solutions. Namely, $\Delta_{\mathrm{th}}^{(ab)}$ is the boundary point of
the regions $a$ and $b$, $\Delta_{\mathrm{th}}^{(bc)}$ is the boundary point
of the regions $b$ and $c$, and $\Delta_{\mathrm{th}}^{(cd)}$ is the
boundary point of the regions $c$ and $d$.}
\label{fig2}
\end{figure}

\subsection{Steady-state positions of the moveable mirror: numerical results}

Let us first analyze the atom-cavity detuning $\Delta _{p,s}$ and the
equilibrium positions $\left\langle x\right\rangle _{s}$ of the mirror. In
principle, we can obtain $\Delta _{p,s}$ by solving the fifth-order equation
in Eq.~(\ref{e}). However, $\Delta _{p,s}$ can also be obtained by plotting
the left $Y_{\mathrm{L}}\!(\Delta _{p,s})$ and right $Y_{\mathrm{R}%
}\!(\Delta _{p,s})$ hand sides of Eq.~(\ref{e}) as functions of $\Delta
_{p,s}$, respectively. The real roots of Eq.~(\ref{e}) can be represented by
the crossing points of two curves corresponding to $Y_{\mathrm{L}}\!(\Delta
_{p,s})$ and $Y_{\mathrm{R}}\!(\Delta _{p,s})$. Here, for clarity, these are
shown
\begin{eqnarray}
Y_{\mathrm{L}}\!\left( \Delta _{p,s}\right) &=&\left\vert 1-\frac{g^{2}N\,%
\tilde{\Omega}\!\left( \Delta _{p,s}\right) }{G\!\left( \Delta _{p,s}\right)
\,\tilde{\Omega}\!\left( \Delta _{p,s}\right) +\frac{\gamma _{0}}{2}\Omega
^{2}}\right\vert ^{2},\;\;\;\; \\
Y_{\mathrm{R}}\!\left( \Delta _{p,s}\right) &=&\frac{\gamma _{0}\kappa }{4%
\tilde{\alpha}_{\mathrm{in}}^{2}\omega _{0}^{2}}\left( \Delta _{p,s}-\Delta
_{0}\right) .
\end{eqnarray}

The curve and the lines, corresponding to $Y_{\mathrm{L}}\!(\Delta _{p,s})$
and $Y_{\mathrm{R}}\!(\Delta _{p,s})$, are schematically shown in Fig.~\ref%
{fig2}. This diagram can be used to analyze the solution of the
self-consistent equation in Eq.~(\ref{e}). In Fig.~\ref{fig2}(a), the
double-well like curve shows how $Y_{\mathrm{L}}\!(\Delta _{p,s})$ changes
with $\Delta _{p,s}$, and the straight line shows how $Y_{\mathrm{R}%
}\!(\Delta _{p,s})$ changes with $\Delta _{p,s}$ for a given $\kappa $ but
for larger $\Delta _{0}=\omega _{a}-\omega _{0}$. Figure~2(b) shows the
solution of the self-consistent equation in Eq.~(\ref{e}) when $\Delta _{0}$
is in the region inside Fig.~2(a) marked by the red dashed box.

In Fig.~\ref{fig2}, the black curves denote the function $Y_{\mathrm{L}%
}\!(\Delta _{p,s})$. The functions $Y_{\mathrm{R}}\!(\Delta _{p,s})$ are
shown by the lines with different colors. The slope of the lines in
Fig.~2(b) are proportional to $\kappa $. The straight lines in Fig.~\ref%
{fig2}(b) correspond to different values of $\Delta _{0}$. Recall that $%
\Delta _{0}$ is the atom-cavity detuning when the two mirrors are fixed.
Each parameter region for $\Delta _{0}$ can have at most five steady-state
solutions and three stable solutions. This system has eight parameters.
However, during most of this study, we will be varying the atom-cavity
detuning $\Delta _{0}$ (for fixed mirrors), and the elastic constant $\kappa
$.\ It is noted that the different crossing points between the transverse
axis and the different lines corresponding to $Y_{\mathrm{R}}\!(\Delta
_{p,s})$ represent the different detunings $\Delta _{0}$. The regions (%
\textit{a-d}) are shown in the figure. The points of intersection of $Y_{%
\mathrm{L}}\!(\Delta _{p,s})$ and $Y_{\mathrm{R}}\!(\Delta _{p,s})$ show the
solutions of Eq.~(\ref{e}) for the whole $\Delta _{p,s}$ region. However,
physically, it is important to study the solution near the atom-cavity
detuning $\Delta _{0}$, and the unstable solution shown by the yellow dot in
Fig.~\ref{fig2}(a) can be neglected.

The letters \textit{a,b,c,d} at the bottom of Fig.~\ref{fig2}(b) represent
regions with different number of solutions. Below, in our discussions for
the steady-state solutions, we only consider the crossing points between two
curves corresponding to $Y_{\mathrm{L}}(\Delta _{p,s})$ and $Y_{\mathrm{R}%
}(\Delta _{p\mathrm{,}s})$, in the red square in Fig.~\ref{fig2}(a). Inside
this red square, the number of real roots of Eq.~(\ref{e}) can be
characterized by three critical values of the detuning $\Delta _{0}$: $%
\Delta _{\mathrm{th}}^{(cd)}$, $\Delta _{\mathrm{th}}^{(bc)}$ and $\Delta _{%
\mathrm{th}}^{(ab)}$, when $\kappa _{\mathrm{L}}<\kappa <\infty $. When
fixing the other parameters, the lower bound $\kappa _{\mathrm{L}}$ of $%
\kappa $ can be fixed as shown in the numerical calculations below.

As schematically shown in Fig.~\ref{fig2}(b), there are four regions in the
space of \textquotedblleft roots\textquotedblright\ divided by the above
three boundaries: (a) when the atom-cavity detuning $\Delta _{0}>\Delta _{%
\mathrm{th}}^{(cd)}$, there is no real root; (b) when $\Delta _{\mathrm{th}%
}^{(bc)}<\Delta _{0}<\Delta _{\mathrm{th}}^{(cd)}$, there always exist two
roots; (c) when $\Delta _{\mathrm{th}}^{(ab)}<\Delta _{0}<\Delta _{\mathrm{th%
}}^{(bc)}$, there are four real roots; (d) when $\Delta _{0}<\Delta _{%
\mathrm{th}}^{(ab)}$, there are two real roots. Also at the threshold points
for $\Delta _{0}=\Delta _{\mathrm{th}}^{(cd)}$, $\Delta _{\mathrm{th}%
}^{(bc)} $, and $\Delta _{\mathrm{th}}^{(ab)}$, the number of steady-state
roots is one, three, and three, respectively. However, in the case when $%
\kappa <\kappa _{\mathrm{L}}$, the threshold value $\Delta _{\mathrm{th}%
}^{(ab)}$ does not exist, and there are only two threshold values, $\Delta _{%
\mathrm{th}}^{(cd)}$ and $\Delta _{\mathrm{th}}^{(bc)}$, which divide the $%
\Delta _{0}$-parameter space into three regions for the roots of Eq.~(\ref{e}%
). In this case, the number of roots will be explained below for given sets
of parameters.

\begin{figure}[tbp]
\includegraphics[bb=7 185 595 638, width=9 cm, height=8 cm, clip]{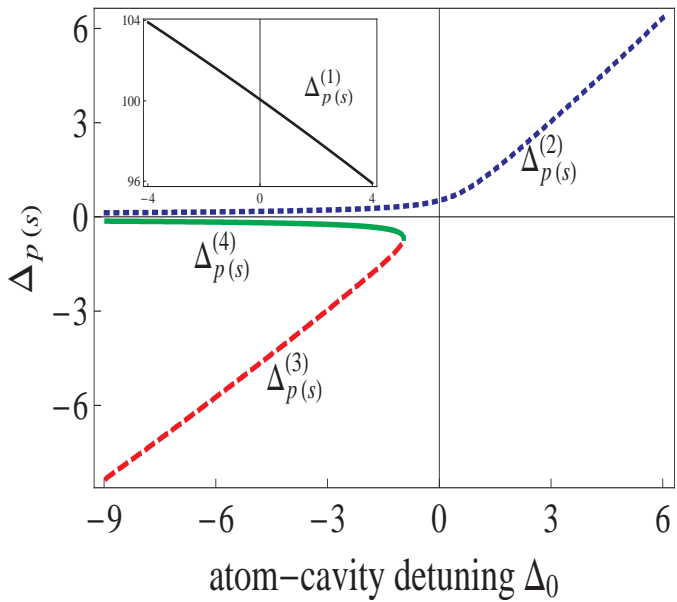}
\caption{(Color Online) The steady state solutions $\Delta _{p,s}^{(i)}$ ($%
i=1,\,2,\,3,\,4$) versus the atom-cavity detunings $\Delta_{0}$ (when the
two mirrors are fixed) for given parameters, e.g., $\protect\kappa =10^{2}$,
$\Delta _{c}=0$, $\protect\omega _{a}=10^{6}$, $\protect\gamma _{0}=10^{-6}$%
, $\protect\gamma _{2}=10^{-4}$, $g=10^{2}$, $\Omega =2$, and $a_{\mathrm{in}%
}=10$. Hereafter, all the quantities are measured in units of $\protect%
\gamma _{1}$. Note that here eight parameters determine the system. Recall
that $\Delta _{p\mathrm{,s}}$ is the steady-state atom-cavity detuning when
one mirror is moveable. The inset shows the steady-state solution $\Delta
_{p,s}^{(1)}$, which corresponds to a very large displacement of the mirror.
As schematically shown in Fig.~\protect\ref{fig2}(b), the dotted blue,
dashed red, and continuous green lines denote the solutions $\Delta
_{p,s}^{(2)}$, $\Delta _{p,s}^{(3)}$ , and $\Delta _{p,s}^{(4)}$,
respectively. For example, the green cross in Fig.~2(b) corresponds to a
single value of $\Delta_{0}$. Here, $\Delta_{0}$ is swept, and the green
cross in Fig.~2(b) becomes a continuous curve. Same for one red dot and one
blue square in Fig.~2(b).}
\label{fig3}
\end{figure}

\begin{figure}[tbp]
\includegraphics[bb=2 212 595 627, width=9 cm, height=8 cm, clip]{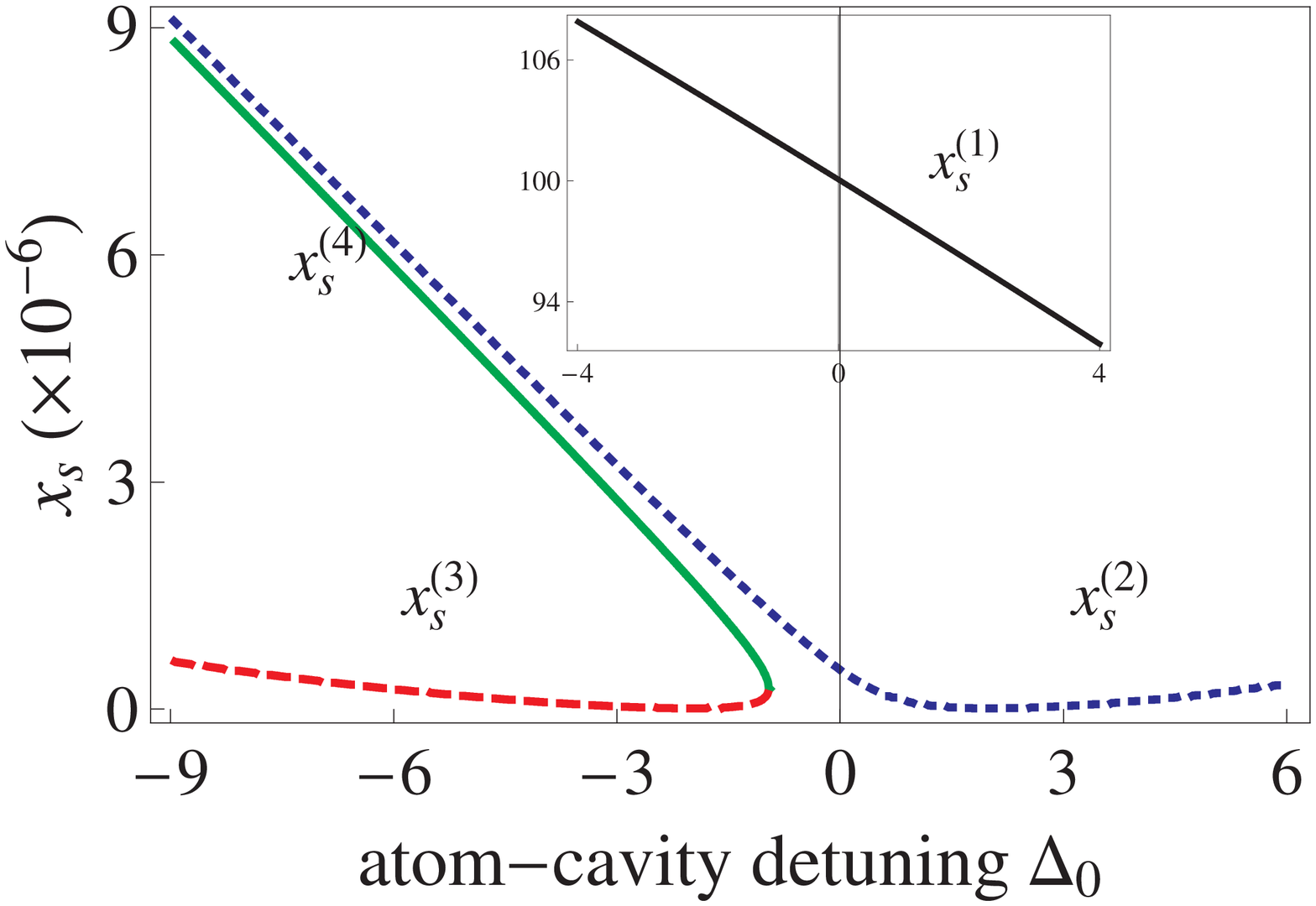}
\caption{(Color Online) Steady-state positions $\langle x\rangle_{s}^{(i)}$ $%
(i=1,\,2,\,3,\, 4)$ for the moveable mirror versus the atom-cavity detuning $%
\Delta_{0}$, for the same parameters listed in Fig.~\protect\ref{fig3}.
Here, the continuous black, dotted blue, dashed red, continuous green lines
denote $\langle x\rangle_{s}^{(1)}$, $\langle x\rangle_{s}^{(2)}$, $\langle
x\rangle_{s}^{(3)}$, and $\langle x\rangle_{s}^{(4)}$, when the atom-cavity
detunings $\Delta _{p,s}$ take the following steady-state values $\Delta
_{p,s}^{(1)}$, $\Delta_{p,s}^{(2)}$, $\Delta_{p\mathrm{,s}}^{(3)}$, and $%
\Delta_{p,s}^{(4)}$, respectively, as shown in Fig.~\protect\ref{fig3}.}
\label{fig4}
\end{figure}

We now simulate the four steady state solutions
\begin{equation*}
\Delta _{p,s}^{(i)}\equiv \Delta _{\text{\textrm{probe,} \textrm{steady-state%
}}}^{(\text{\textrm{particular solution label}})}\text{, }(i=1,\,2,\,3,\,4)
\end{equation*}%
for given parameters as in Ref.~\cite{liyong}, e.g., $\Delta _{c}=0$, $%
\omega _{a}=10^{6}$, $\gamma _{0}=10^{-6}$, $\gamma _{2}=10^{-4}$, $g\sqrt{N}%
=10^{2}$, $\Omega =2$ and $a_{\mathrm{in}}=10$. Hereafter, all quantities
are measured in units of $\gamma _{1}$. With the above parameters, we can
find that the lower bound $\kappa _{\mathrm{L}}$ is about $6400$.

We first study the case for $\kappa <\kappa _{\mathrm{L}}\cong 6400$, e.g., $%
\kappa =10^{2}$. In this case, there are two critical values $\Delta _{%
\mathrm{th}}^{(a)}\cong 25$ and $\Delta _{\mathrm{th}}^{(b)}\cong -0.95$.
Fig.~3 shows the steady-state atom-cavity detunings $\Delta _{p,s}^{(i)}$
(when the mirror moves) versus $\Delta _{0}$ (when the mirror is fixed). We
find that there is no solution when $\Delta _{0}=\omega _{a}-\omega
_{0}\gtrsim 25$. This means that when the difference between the atomic
frequency $\omega _{a}$ and the cavity frequency $\omega _{0}$ is larger
than $25$, there is no steady-state near the detuning $\Delta _{c}$. When $%
-0.95\lesssim \Delta _{0}\lesssim 25$, there are two steady-state solutions,
i.e., $\Delta _{p,s}^{(1)}$ and $\Delta _{p,s}^{(2)}$ shown in Fig.~\ref%
{fig3}. In this region, we find that $\Delta _{p,s}^{(1)}$ decreases
linearly, but $\Delta _{p,s}^{(2)}$ increases with increasing $\Delta _{0}$.
When $\Delta _{0}\lesssim -0.95$, as shown in Fig.~\ref{fig3}, there are
four steady-state solutions, i.e., $\Delta _{p,s}^{(i)}$ ($i=1,\,2,\,3,\,4$%
). Figure~\ref{fig3} shows that when the detuning $\Delta _{0}$ passes
through $-0.95$, from the right to the left, two additional solutions ($%
\Delta _{p,s}^{(3)}$ and $\Delta _{p,s}^{(4)}$) appear. We also find that
two solutions ($\Delta _{p,s}^{(2)}$ and $\Delta _{p,s}^{(4)}$) gradually
approach $\Delta _{c}=0$ to realize the two-photon-resonance condition.
Moreover, note that $\Delta _{p,s}^{(3)}$ increases almost linearly with
increasing $\Delta _{0}$.

With the same parameters as those in Fig.~3, we have also plotted in Fig.~%
\ref{fig4} the $\Delta _{0}$-dependent location $\left\langle x\right\rangle
_{s}^{(i)}$ of the mirror corresponding to $\Delta _{p,s}^{(i)}$. We find
that there is no solution for the steady-state value of $\left\langle
x\right\rangle _{s}^{(i)}$ when $\Delta _{0}\gtrsim 25$. When $-0.95\lesssim
\Delta _{0}\lesssim 25$, the mirror's position $\left\langle x\right\rangle
_{s}^{(1)}$, corresponding to the solution $\Delta _{p,s}^{(1)}$, exhibits a
very large (compared with $\left\langle x\right\rangle _{s}^{(2)}$, $%
\left\langle x\right\rangle _{s}^{(3)}$, $\left\langle x\right\rangle
_{s}^{(4)}$) displacement of the mirror, and $\left\langle x\right\rangle
_{s}^{(1)}$ increases as $\Delta _{0}$ increases. The mirror's displacement $%
\left\langle x\right\rangle _{s}^{(2)}$, corresponding to the solution $%
\Delta _{p,s}^{(2)}$, is nearly zero. When $\Delta _{0}\lesssim -0.95$, the
four steady-state solutions for the displacement $\left\langle
x\right\rangle _{s}$ exist simultaneously. Two of them, i.e., $\left\langle
x\right\rangle _{s}^{(2)}$ and $\left\langle x\right\rangle _{s}^{(4)}$,
show that the spring is compressed, and their displacements linearly
increase when $\Delta _{0}$ decreases. One of them, i.e., $\left\langle
x\right\rangle _{s}^{(3)}$, nearly vanishes.

From Eq.~(\ref{eq:15}), we know that the oscillating mirror can affect the
two-photon resonance by changing the effective frequency $\omega _{L}$ of
the cavity field. Because when the mirror is fixed, the two-photon resonant
condition becomes
\begin{equation}
\Delta _{0}=\omega _{a}-\omega _{0}=\Delta _{c}.  \label{eq:29}
\end{equation}%
However, this condition is modified to
\begin{equation}
\Delta _{p,s}=\omega _{a}-\omega _{L}=\Delta _{c}\,,  \label{eq:30}
\end{equation}%
when the mirror is oscillating. We note that the two-photon resonant
condition in Eq.~(\ref{eq:30}) is further modified to
\begin{equation}
\Delta _{p,s}^{(i)}=\omega _{a}-\omega _{L}=\Delta _{c}\,,  \label{eq:31}
\end{equation}%
when the system reaches its steady state. We assume that the two-photon
resonant condition in Eq.~(\ref{eq:29}) is satisfied when the mirror is
fixed. Recall that $\Delta _{c}=0$ when Fig.~\ref{fig3} and Fig.~\ref{fig4}
are plotted. This means that the two-photon resonant condition is $\Delta _{p%
\mathrm{,}s}^{(i)}=0$ in this case. We find that $\left\langle
x\right\rangle _{s}$ is nearly zero as shown in Fig.~\ref{fig4} when $\Delta
_{0}>\Delta _{c}=0$. In this case, there is no value of $\Delta _{p,s}$
approaching zero, as shown in Fig.~\ref{fig3}; so the two-photon resonance
cannot happen. From Fig.~\ref{fig3} and Fig.~\ref{fig4}, we find that the
two-photon resonance $\Delta _{p,s}^{(i)}=\Delta _{c}=0$ might happen when $%
\Delta _{0}\lesssim -0.95$. Because in this region, $\Delta _{p,s}^{(2)}$
and $\Delta _{p,s}^{(4)}$ can approach zero as in Fig.~\ref{fig3}, which
correspond to the steady-state values of the mirror's positions $%
\left\langle x\right\rangle _{s}^{(2)}$ and $\left\langle x\right\rangle
_{s}^{(4)}$, respectively, as in Fig.~\ref{fig4}.

\begin{figure}[tbp]
\includegraphics[bb=7 169 595 614, width=9 cm, height=8 cm, clip]{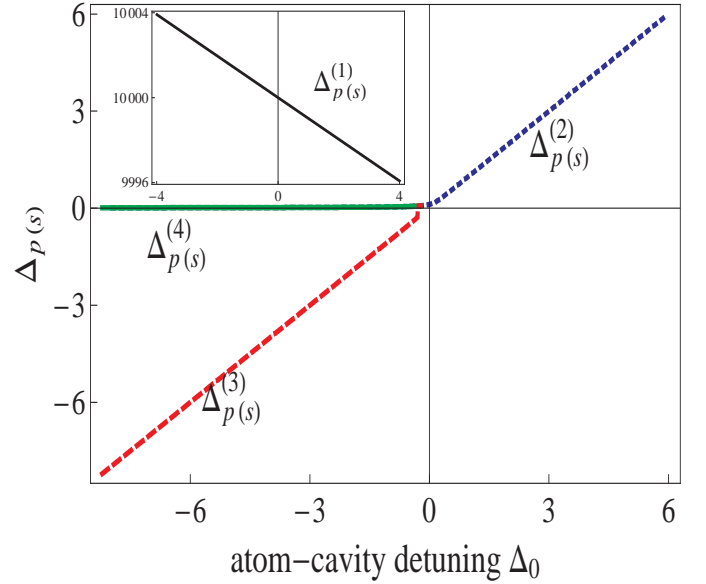}
\caption{(Color Online) Steady-state solutions $\Delta _{p(s)}^{(i)}$ ($%
i=1,\,2,\,3,\,4$) schematically shown in Fig.~\protect\ref{fig2}(b) versus $%
\Delta_{0}$ with $\protect\kappa =10^{4}$. The other parameters here are the
same as in Fig.~\protect\ref{fig3}. The inset shows the steady state
solution $\Delta _{p(s)}^{(1)}$ that corresponds to a very large
displacement of the mirror. As shown in Fig.~\protect\ref{fig2}(b), the
dotted blue, dashed red, and continuous green lines denote the solutions $%
\Delta _{p(s)}^{(2)}$, $\Delta _{p(s)}^{(3)}$ , and $\Delta _{p(s)}^{(4)}$,
respectively.}
\label{fig5}
\end{figure}

\begin{figure}[tbp]
\includegraphics[bb=7 176 595 591, width=9 cm, height=8 cm, clip]{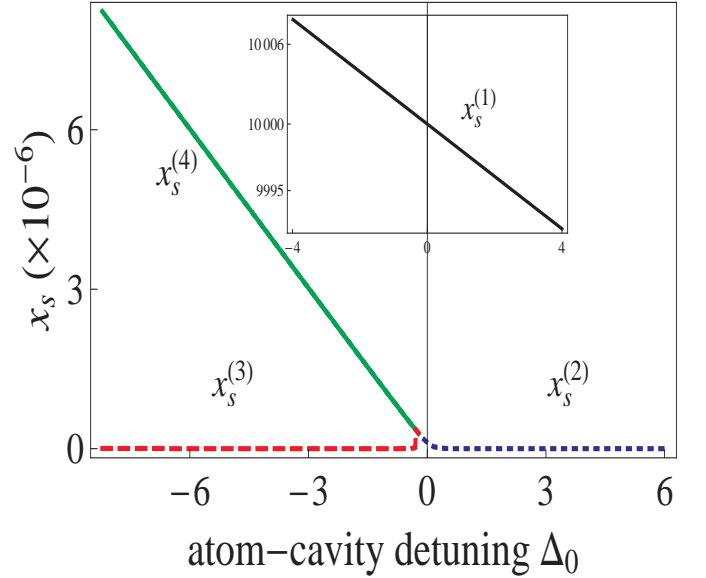}
\caption{(Color Online) Steady-positions $\langle x\rangle_{s}^{(i)}$ $%
(i=1,\,2,\,3,\, 4)$ of the moveable mirror versus the atom-cavity detuning $%
\Delta_{0}$, with the same parameters as in Fig.~\protect\ref{fig5}. Here,
the continuous black, dotted blue, dashed red, continuous green lines denote
$\langle x\rangle_{s}^{(1)}$, $\langle x\rangle_{s}^{(2)}$, $\langle
x\rangle_{s}^{(3)}$, and $\langle x\rangle_{s}^{(4)}$, when the atom-cavity
detunings $\Delta _{p(s)}$ (for a moving mirror) takes the following
steady-state values: $\Delta _{p,s}^{(1)}$, $\Delta_{p(s)}^{(2)}$, $%
\Delta_{p,s}^{(3)}$, and $\Delta_{p,s}^{(4)}$, respectively, as shown in
Fig.~\protect\ref{fig5}.}
\label{fig6}
\end{figure}

We now turn to study the steady-state values of $\Delta _{p,s}$ and $%
\left\langle x\right\rangle _{s}$ for the case when $6400\lesssim \kappa
<\infty $, e.g., $\kappa =10^{4}$. In this case, as schematically shown in
Fig.~\ref{fig2}(b), there are three critical values $\Delta _{\mathrm{th}%
}^{(i)}$ ($i=ab,\,bc,\,cd$): $\Delta _{\mathrm{th}}^{(cd)}\cong 2.5\times
10^{3}$, $\Delta _{\mathrm{th}}^{(bc)}\cong -0.22$ and $\Delta _{\mathrm{th}%
}^{(ab)}\cong -2.5\times 10^{5}$. The number of solutions for $\Delta _{p,s}$
has the same descriptions as for Fig.~\ref{fig2}(b). Similarly to Fig.~\ref%
{fig3} and Fig.~\ref{fig4}, we plot Fig.~\ref{fig5} and Fig.~\ref{fig6} for $%
\Delta _{p,s}^{(i)}$ and their corresponding $\left\langle x\right\rangle
_{s}^{(i)}$, respectively. The two-photon resonance condition in Eq.~(\ref%
{eq:31}) for $\Delta _{c}=0$ might also be satisfied in this case. Because
when $\Delta _{0}\lesssim -0.22$, two steady-state values ($\Delta
_{p,s}^{(2)}$ and $\Delta _{p,s}^{(4)}$) of $\Delta _{p,s}$ are near zero,
as shown in Fig.~\ref{fig5}, which correspond to the steady-state solution
of the mirror's position $\left\langle x\right\rangle _{s}^{(2)}$ and $%
\left\langle x\right\rangle _{s}^{(4)}$, as shown in Fig.~\ref{fig6}. It is
also found that one, i.e., $\left\langle x\right\rangle _{s}^{(2)}$, of the
steady-state solutions of $\left\langle x\right\rangle _{s}$ nearly
vanishes, as shown in Fig.~\ref{fig6}, when $\Delta _{0}>\Delta _{c}=0$.

Finally, we note that there is only one steady state solution $\Delta
_{p,s}=\Delta _{0}$ for Eq.~(\ref{e}) in the limit $\kappa \rightarrow
\infty $. This implies that when the elastic coefficient $M\omega _{M}^{2}$
is very large, it is difficult for the photon pressure to make the mirror to
even have a tiny displacement, and the oscillating mirror does not affect
the optomechanical system~\cite{liyong1}.

\section{Electromagnetically Induced transparency in the optomechanical
system}

To explore how the mirror's oscillation affects the EIT, let us now study
the susceptibility of the atomic medium. As discussed above, we consider the
single-mode cavity field as the probe field. In this case, we have the
electric field
\begin{equation}
E(t)=\sqrt{\frac{\omega _{L}}{2V\epsilon _{0}}}ae^{-i\omega _{L}t}+\text{%
\textrm{H.c.}}=\varepsilon e^{-i\omega _{L}t}+\text{\textrm{H.c.}}\,.
\label{a2}
\end{equation}%
The linear response of the atomic ensemble to the weak probe field $E(t)$
can be described by the susceptibility
\begin{equation}
\chi =\frac{\left\langle p\right\rangle }{\left\langle \varepsilon
\right\rangle \epsilon _{0}}.  \label{a3}
\end{equation}%
Here, the average polarization $\left\langle p\right\rangle $ of the atomic
ensemble is
\begin{equation}
\left\langle p\right\rangle =\frac{\mu }{V}\sum_{i=1}^{N}\sigma
_{ba}^{\left( i\right) }.
\end{equation}%
Using Eqs.~(\ref{a1}), (\ref{a2}) and (\ref{a3}), we obtain the
susceptibility $\chi $
\begin{eqnarray}
\chi &=&F\frac{\gamma _{2}\Xi -\left( \Delta _{p,s}-\Delta _{c}\right)
\Theta }{\Xi ^{2}+\Theta ^{2}}  \notag \\
&&+iF\frac{\gamma _{2}\Theta +\left( \Delta _{p,s}-\Delta _{c}\right) \Xi }{%
\Xi ^{2}+\Theta ^{2}},
\end{eqnarray}%
with
\begin{eqnarray}
F &=&\frac{\mu ^{2}N}{\epsilon _{0}V}, \\
\Xi &=&\gamma _{1}\left( \Delta _{p,s}-\Delta _{c}\right) +\gamma _{2}\Delta
_{p,s},
\end{eqnarray}%
(where $F$ is proportional to the density $N/V$) and
\begin{equation}
\Theta =\Omega ^{2}-\Delta _{p,s}\left( \Delta _{p\mathrm{,}s}-\Delta
_{c}\right) +\gamma _{1}\gamma _{2}.
\end{equation}%
\begin{figure}[tbp]
\includegraphics[bb=6 5 360 234, width=9 cm, height=8 cm, clip]{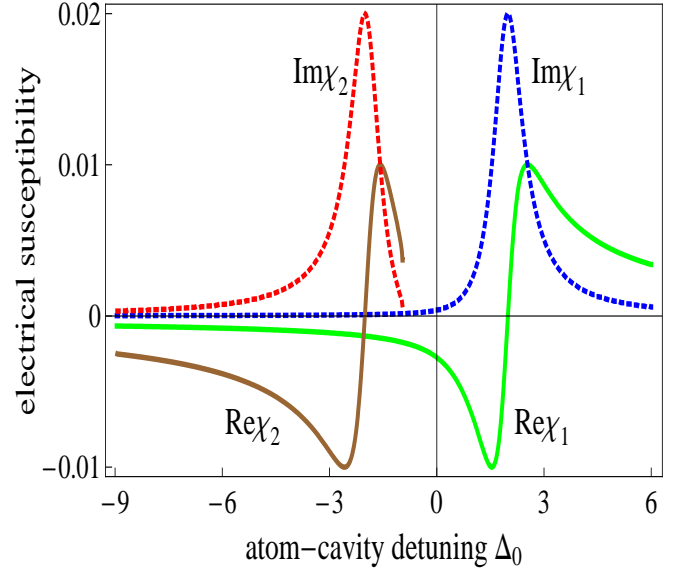}
\caption{(Color Online) The susceptibilities $\protect\chi _{1}$ and $%
\protect\chi _{2}$ for the detuning between the atom and the control field $%
\Delta _{c}=0$ and $\protect\kappa =10^{2}$. We plot only two
susceptibilities because there are three stable solutions in this parameter
region but two of them are quite similar. Here, the dotted and solid curves
correspond to the imaginary and real parts, respectively. The blue and the
green colors correspond to $\protect\chi _{1}$ and the dark red and brown
colors correspond to $\protect\chi _{2}$, respectively. The EIT-like region
is located between the two peaks of Im$\protect\chi _{1}$ and Im$\protect%
\chi _{2}$.}
\label{fig7}
\end{figure}
\begin{figure}[tbp]
\includegraphics[bb=6 5 326 213, width=9 cm, height=8 cm, clip]{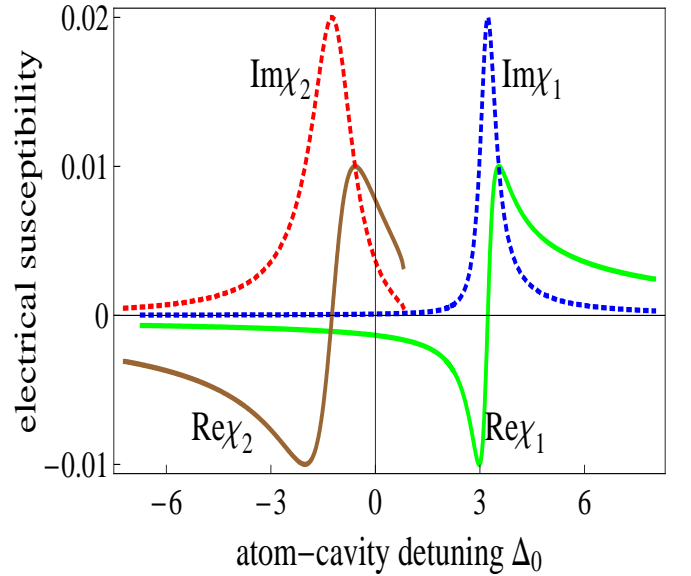}
\caption{(Color Online) The susceptibilities $\protect\chi _{1}$ and $%
\protect\chi _{2}$ for $\Delta _{c}=2$ and $\protect\kappa =10^{2}$. Here,
the dotted and solid curves correspond to the imaginary and real parts,
respectively. The blue and the green colors correspond to $\protect\chi _{1}$
and the dark red and brown colors correspond to $\protect\chi _{2}$,
respectively. The EIT-like region is located between the two peaks of Im$%
\protect\chi _{1}$ and Im$\protect\chi _{2}$.}
\label{fig8}
\end{figure}
It is well known that the real, $\mathrm{Re}(\chi )$, and imaginary, $%
\mathrm{Im}(\chi )$, parts of the susceptibility $\chi $ describe the
dispersion and absorption of light, respectively.

As discussed in the last section, one of the solutions corresponding to $%
x_{s}^{(1)}$ and $\Delta _{p,s}^{(1)}$ is unstable, and another solution,
corresponding to $x_{s}^{(4)}$ and $\Delta _{p\mathrm{,}s}^{(4)}$, is
similar to the solution $x_{s}^{(2)}$ and $\Delta _{p,s}^{(2)}$ in the
region ($\Delta _{\mathrm{th}}^{(b)}$, $\Delta _{\mathrm{th}}^{(c)}$).
Therefore, we only consider the other two solutions below. In Fig.~\ref{fig7}%
, $\mathrm{Re}(\chi _{i})$ and $\mathrm{Im}(\chi _{i})$ ($i=1,\,2$) versus $%
\Delta _{0}$ are plotted for the two solutions $\Delta _{p,s}^{(2)}$ and $%
\Delta _{p,s}^{(3)}$ studied in Fig.~3. Here, $\chi _{1}$ and $\chi _{2}$
denote the susceptibilities corresponding to $\Delta _{p,s}^{(2)}$ and $%
\Delta _{p,s}^{(3)}$, respectively. All parameters in Fig.~7 are the same as
those in Fig.~\ref{fig3}. It is found that when $\Delta _{0}>\Delta _{c}=0$,
the curves for $\mathrm{Re}(\chi _{1})$ and $\mathrm{Im}(\chi _{1})$ are
similar to those in the usual EIT phenomenon~\cite{liyong1}. This means that
the steady state value $\left\langle x\right\rangle _{s}^{(2)}$ of the
mirror's displacement is near zero in this region of parameters, and the
oscillating mirror has no effect on the EIT. However, when $\Delta
_{0}<\Delta _{c}=0$, the mirror's displacement $\left\langle x\right\rangle
_{s}^{(2)}$ makes the two-photon-resonance condition $\Delta
_{p,s}^{(2)}=\omega _{a}-\omega _{L}\approx \Delta _{c}=0$ be approximately
satisfied. As a result, in this region, each of the curves $\mathrm{Re}(\chi
_{1})$ and $\mathrm{Im}(\chi _{1})$ is almost close to zero, like an
infinite \textquotedblleft tail". When $\Delta _{0}\lesssim -0.95$ as shown
in Fig.~\ref{fig3}, another solution $\Delta _{p\mathrm{,}s}^{(3)}$ emerges,
thus we also have a $\Delta _{p,s}^{(3)}$-dependent susceptibility $\chi
_{2} $.

As shown in Fig.~\ref{fig7}, the curves corresponding to the real $\mathrm{Re%
}(\chi _{2})$ and imaginary $\mathrm{Im}(\chi _{2})$ parts of the
susceptibility $\chi _{2}$ are similar to those in the usual EIT~\cite%
{liyong1}, since the mirror's displacement $\left\langle x\right\rangle
_{s}^{(3)}$ is nearly zero. From Fig.~\ref{fig7}, we find that the right
part of the curve $\mathrm{Im}(\chi _{2})$ almost merges with the left part
of the curve Im$\chi _{1}$, so that a \textit{transparency}
\textquotedblleft window" is formed.

\begin{figure}[tbp]
\includegraphics[bb=6 5 335 220, width=9 cm, height=8 cm, clip]{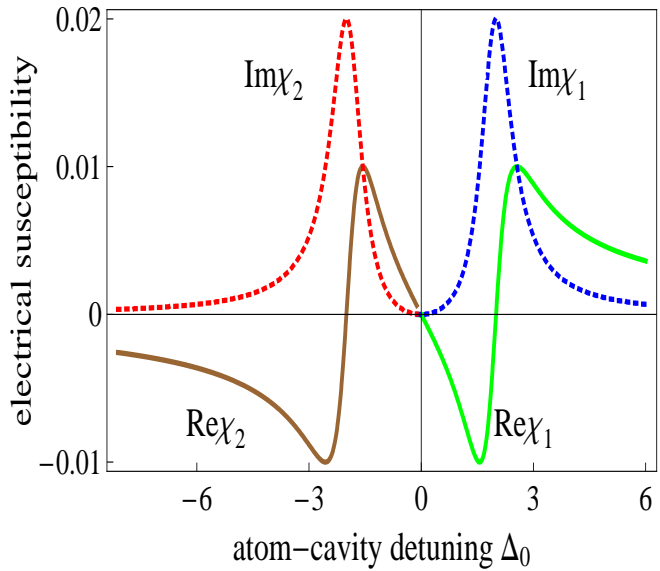}
\caption{(Color Online) The susceptibilities for $\Delta _{c}=0$ and $%
\protect\kappa =1.6\times10^{10}$. Note that this value of the elastic
constant $\protect\kappa$ is huge, corresponding to an almost fixed mirror.
In this case, the Re$\protect\chi _{1}$ and Im$\protect\chi _{1}$ occur for
mostly positive values of the atom-cavity detuning. When the spring constant
becomes softer, as in Fig.~7, the Re$\protect\chi _{1}$ and Im$\protect\chi%
_{1}$ have a response that extend over a huge range of values of the
atom-cavity detuning $\Delta _{0}$, even for $\Delta _{c}\gtrsim -6$. When $%
\protect\kappa $ tends to infinity, the susceptibility becomes the same as
that in the usual EIT phenomenon. This consistency check is reassuring for
our calculations. Here, the dotted and solid curves correspond to the
imaginary and real parts, respectively. The blue and the green colors
correspond to $\protect\chi _{1}$ and the dark red and brown colors
correspond to $\protect\chi _{2}$, respectively. The EIT-like region occurs
near zero detuning $\Delta _{0}$.}
\label{fig9}
\end{figure}

To know how $\Delta _{c}$ affects the EIT, we can also study the
susceptibility $\chi $ for the detuning $\Delta _{c}=2$ when other
parameters are assumed to be the same as those in Fig.~\ref{fig7}. In this
case, the steady-state solutions of $\Delta _{p,s}$ and $\left\langle
x\right\rangle _{s}$ are similar to those for $\Delta _{c}=0$. The curves
corresponding to these solutions are just rightward shifts for the curves in
Fig.~\ref{fig3} and Fig.~\ref{fig4}, but the shapes of the curves are almost
the same. Similar to Fig.~\ref{fig7}, we choose two steady-state solutions
and plot $\mathrm{Re}(\chi _{i})$ and $\mathrm{Im}(\chi _{i})$ ($i=1,\,2$)
versus $\Delta _{0}$. Obviously, the essential conclusions remain unchanged
as those in Fig.~\ref{fig7}, but all curves have a rightward shift.

Based on the analysis in Sec.~III, $\Delta _{\mathrm{th}}^{(bc)}$ and $%
\Delta _{\mathrm{th}}^{(ab)}$ approach zero from the left side when $\kappa $
is increased. When $\kappa >\kappa _{L}$, e.g., $\kappa \gtrsim 6400$ in
Fig.~\ref{fig5} and Fig.~\ref{fig6}, the larger $\kappa $ corresponds to
shorter \textquotedblleft tails" of the curves corresponding to $\mathrm{Re}%
(\chi _{1})$ and $\mathrm{Im}(\chi _{1})$ when $\Delta _{0}\gtrsim -0.95$.
In the limit $\kappa \rightarrow \infty $, the \textquotedblleft tails" (for
$\Delta _{0}\gtrsim 0$) disappear and the right part of the curve $\mathrm{Im%
}(\chi _{2})$ just meets the left part of the curve $\mathrm{Im}(\chi _{1})$
to form a \textit{transparency} \textquotedblleft window". In this limit,
all the physical properties return to that in the usual EIT phenomenon.
Namely, we recover the standard EIT when $\kappa \rightarrow \infty $. This
asymptotic result is shown in Fig.~\ref{fig9} with a large $\kappa $ (e.g., $%
\kappa =1.6\times 10^{10}$), but other parameters are the same as in Fig.~%
\ref{fig3}. In Fig.~\ref{fig9}, around the point $\Delta _{0}>\Delta _{c}=0$%
, the left parts of the curves $\mathrm{Re}(\chi _{1})$ and $\mathrm{Im}%
(\chi _{1})$ nearly merge with the right parts of the curves $\mathrm{Re}%
(\chi _{2})$ and $\mathrm{Im}(\chi _{2})$, and thus transparency windows are
formed as in the usual EIT~\cite{liyong1}.

\section{Conclusion and Remarks}

We have studied the effects of the end mirror's oscillation on the EIT
phenomenon for an atomic ensemble confined in a gas cell placed in a
micro-cavity. This study can help us to quantitatively consider the quantum
interface between an optomechanical system and an atomic gas displaying EIT.
The results obtained could be used to improve the measurement precision
based on the EIT effect, when the medium is placed inside a microcavity with
a one-end oscillating mirror and the cavity acts as the probe light. We have
shown that the whole system exhibits multi-stability due to the mirror's
vibration, and we have also numerically obtained the threshold values of the
parameters, which can help determine how many steady-states exist in the
corresponding parameter regions. This multi-stability can be explicitly
displayed through a modified EIT phenomenon. Consequently, we investigate
the effects of the multi-steady-state solutions on the EIT phenomenon and
find that in some parameter regions there are two solutions that
approximately satisfy the two-photon resonance condition. Therefore, the
properties of both the real and imaginary parts of the susceptibility are
significantly altered. When the spring elastic constant $\kappa $ increases,
the mirror becomes less moveable, and in this case all properties of the
system gradually revert to those of the usual EIT phenomenon.

\acknowledgments

C. P. Sun acknowledges supports by the NSFC with Grants No. 10474104, No.
60433050, and No. 10704023, NFRPCNo. 2006CB921205 and 2005CB724508. F. Nori
was supported in part by the National Security Agency (NSA), the Laboratory
for Physical Sciences (LPS), the Army Research Office (ARO), the National
Science Foundation (NSF) grant No. EIA-0130383, and the JSPS-RFBR
06-02-91200.


\end{document}